\documentstyle[eqsecnum,prl,aps,epsf]{revtex}
\begin{document}
\tighten
\twocolumn[\hsize\textwidth\columnwidth\hsize\csname
@twocolumnfalse\endcsname
\preprint{WISC-MILW-00-TH-2}
\draft
\title{A power filter for the detection of burst sources of gravitational 
radiation in interferometric detectors} 
\author{Warren G. Anderson, Patrick R. Brady, Jolien D. E. Creighton}
\address{Department of Physics, University of Wisconsin - Milwaukee, 
P.O. Box 413, Wisconsin, 53201, U.S.A.}
\author{\'{E}anna \'{E}. Flanagan}
\address{Newman Laboratory, Cornell University Ithaca, New York 14853-5001, 
U.S.A.}

\maketitle
\begin{abstract}%
We present a filter for detecting gravitational wave signals from burst
sources. This filter requires only minimal advance knowledge of the 
expected signal: i.e. the signal's frequency band and time duration. 
It consists of a threshold on the total power in the data stream in 
the specified signal band during the specified time. This filter is optimal
(in the Neyman-Pearson sense) for signal searches where only this minimal
information is available.
\end{abstract}

\narrowtext
\twocolumn
\vskip0pc]

\section{Introduction}	%) A SECTION HEADING
\noindent
Currently, the best understood and most highly developed technique for
detecting gravitational waves with interferometric detectors is matched
filtering. This technique is optimal if the waveform to be detected is
known in advance. There are, however, potentially important sources of
gravitational radiation that are not well enough modeled to obtain reliable
waveforms. Included in this category are binary black hole mergers, which have
been discussed in some detail by Flanagan and Hughes\cite{Flanagan_E:1998}
(FH), and supernovae\cite{Thorne_K:1995}.

In order to detect poorly modeled sources, new techniques must be developed. 
Clearly, these techniques must perform well with incomplete prior knowledge 
of the expected signal. A number of techniques are under active 
investigation\cite{filters}. 

The purpose of this article is to discuss one such technique, the power
filter\cite{Flanagan_E:1998,pow_filt}. This filter only requires prior 
knowledge of the duration and frequency band of the signal. It is therefore 
well suited to detecting black hole merger signals, for which FH have estimated
these parameters. Furthermore, we shall show in the following sections that 
this filter is optimal in the sense that it gives the highest probability of
correctly detecting a signal for a given false alarm probability. Our 
treatment here is cursory; a more comprehensive description of the filter
is in preparation\cite{pow_filt}. 

\section{The Power Statistic}
\noindent
Consider the output $h(t)$ of the interferometric gravitational wave detector. 
It is sampled at a finite rate $1/\Delta t$ to produce a time series
$h_j=h(j\Delta t)$, where $j$ is an integer. A segment of $N$ samples defines
a vector ${\mathbf{h}}=(h_j,\ldots,h_{(j+N-1)})$. This vector can be written 
as
\begin{equation}
\label{e:output}
  \mathbf{h} = \mathbf{n} + \mathbf{s}
\end{equation}
where $\mathbf{n}$ is detector noise and $\mathbf{s}$ is a (possibly absent) 
signal. 

The noise is assumed to be stochastic. The vectors $\mathbf{n}$ and 
$\mathbf{h}$ are therefore described statistically. Statistical fluctuations 
lead to two types of errors in detecting a signal: false alarms, in which
signals are detected when none are present, and false dismissals, in which
signals are not detected when present. An optimal filter is defined to be one 
which minimizes false dismissals for a given false alarm rate.

Neyman and Pearson have shown that an optimal filter is one for which the 
likelihood ratio $\Lambda$ is compared to a 
threshold\cite{likelihood,Kassam_S:1988}. The likelihood ratio is defined to 
be
\begin{equation}
   \Lambda[{\mathbf{h}}]\equiv \int {\cal D}[{\mathbf{s}}]~
      \frac{p[\mathbf{h}|\mathbf{s}]}{p[\mathbf{h}|\mathbf{0}]},
\label{e:likelihood_ratio}
\end{equation}
where $p[\mathbf{h}|\mathbf{s}]$ ($p[\mathbf{h}|\mathbf{0}]$) is the 
probability of obtaining $\mathbf{h}$ given that a signal $\mathbf{s}$ is
present (absent) and ${\cal D}[\mathbf{s}]$ is a measure on the space of 
signals.

The quantities $p[\mathbf{h}|\mathbf{s}]$ and $p[\mathbf{h}|\mathbf{0}]$
depend on the statistical properties of the noise. For convenience, we 
assume in this article that the noise is stationary and Gaussian\footnote{
However, other types of noise can also be considered\cite{pow_filt}.}. We can 
therefore write the probability distribution for the noise as
\begin{equation}
   p[{\mathbf{n}}]= C \exp\left[-\frac{\mathbf{n}\cdot \mathbf{n}}{2} \right],
\label{e:noise_dist}
\end{equation}
where $C$ is a constant prefactor and $\mathbf{n}\cdot\mathbf{n}$ an inner 
product, both of which are determined by the autocorrellation matrix of the 
noise. When a signal is present (absent) we have 
$\mathbf{n}=\mathbf{h}-\mathbf{s}$ ($\mathbf{n}=\mathbf{h}$), and can easily 
use Eq. (\ref{e:noise_dist}) to find the integrand in 
Eq. (\ref{e:likelihood_ratio})
\begin{equation}
   \frac{p[{\mathbf{h}\mid\mathbf{s}}]}{p[{\mathbf{h}\mid\mathbf{0}}]} =   
      \exp\left[({\mathbf{s}\cdot\mathbf{h}})-\frac{1}{2} 
      (\mathbf{s}\cdot\mathbf{s})\right]. 
\label{e:1}
\end{equation}
The measure ${\cal D}[\mathbf{s}]$ in Eq. (\ref{e:likelihood_ratio}) reflects
our prior knowledge about the signal. For those signal parameters about which
we have no prior knowledge, we choose a measure which reflects our ignorance.

Consider now the case where we know the time window and frequency band in
which a signal occurs. The measure ${\cal D}[\mathbf{s}]$ restricts the
integral in Eq.  (\ref{e:likelihood_ratio}) to the projection
${\mathbf{h_\parallel}}$ of $\mathbf{h}$ into the space of vectors with the
give window and frequency band. Introducing the notation
$A^2=\mathbf{s}\cdot\mathbf{s}$, ${\cal E}=
\mathbf{h}_\parallel\cdot\mathbf{h}_\parallel$ and
${\mathbf{s}\cdot\mathbf{h}}_\parallel=A {\cal E}^{1/2} \cos\theta$, we rewrite
Eq. (\ref{e:likelihood_ratio}) as
\begin{equation}
   \Lambda[{\mathbf{h}}] = \int {\cal D}[\theta,A]~ \exp\left[A {\cal E}^{1/2}
      \cos\theta-\frac{1}{2} A^2 \right].
\label{e:2}
\end{equation}
Since we claim no prior knowledge of $\theta$, a suitable measure over 
$\theta$ is uniform over all possible angles between ${\mathbf{h}}_\parallel$ 
and ${\mathbf{s}}$ (i.e. over an ${\cal N}$ sphere, where ${\cal N}$ is the
dimension of the space of vectors with the required duration and frequency
band), which reflects our lack of knowledge. 

While a suitable measure over the signal amplitude $A$ can also be deduced, it
is unnecessary here. Instead, one constructs a locally optimal 
statistic\cite{Kassam_S:1988}, $\Lambda_{\mathrm{loc}}[\mathbf{h}]$, which is
appropriate in the limit of weak signals. To construct this statistic, one
expands the likelihood ratio (\ref{e:2}) in a Taylor series about $A=0$. The
statistic is simply the first non-vanishing coefficient (excluding the $A^0$ 
coefficient) in the expansion. Expanding (\ref{e:2}) and integrating over 
$\theta$ we get
\begin{equation}
   \Lambda_{\mathrm{loc}}[{\mathbf{h}}]\propto\ {\cal E} + \mbox{terms
      independent of $\mathbf{h}$}.
\label{eq:local_stat}
\end{equation}
The terms independent of ${\mathbf{h}}$ clearly do not discriminate between
the presence and absence of a signal. Thus, the optimal statistic for
detecting a signal of known duration and frequency band is simply the total
power in the detector output over that time and band. 

\section{Operating Characteristics}
\noindent
In the previous section we determined the optimal statistic ${\cal E}$ for
signals of known duration and frequency band in stationary Gaussian noise.  We
construct the optimal filter from this statistic via a threshold decision
rule. That is, we calculate at what value ${\cal E}^\star$ of the statistic we
incur the largest acceptable false alarm probability. We then compare values
of ${\cal E}$ calculated from our data with ${\cal E}^\star$. A signal is said
to have been detected if ${\cal E} > {\cal E}^\star$. 

For Gaussian noise, false alarm and false dismissal probabilities can be 
calculated analytically up to quadrature. If no signal is present, 
${\cal E}$ is just the sum of the squares of 
$V\equiv2\times\delta t\times \delta f$ independent Gaussian random variables. 
Thus ${\cal E}$ has a $\chi^2$ distribution with $V$ degrees of freedom, and 
the false alarm probability for a value ${\cal E}^\star$ is just
\begin{equation}
\label{e:falseAlarm}
  P({\cal E}>{\cal E}^\star \mid A=0) = 
   \frac{\Gamma(V/2,{\cal E}^\star/2)}{\Gamma(V/2)}
\end{equation}
where $\Gamma(a,x)=\int_x^\infty e^{-t}t^{a-1}dt$ is the incomplete Gamma
function.  

If a signal of amplitude $A$ is present, ${\cal E}$ is distributed as a 
weighted sum of $\chi^2$ probability distributions,
\begin{equation}
\label{e:excessPowerProbabilityDistribution}
  p({\cal E} \mid V, A) = \sum_{n=0}^\infty
  \frac{e^{-A^2/2}(A^2/2)^n}{n!}
  \frac{e^{-{{\cal E}/2}}({\cal E}/2)^{n+V/2-1}}{\Gamma(n+V/2)}.
\end{equation}
This is the non-central $\chi^2$ probability distribution. The false dismissal
probability is given by
\begin{equation}
\label{e:trueDetection}
  P({\cal E}<{\cal E}^\star\mid A) = \int^{{\cal E}^\star}_0 
      p({\cal E} \mid V, A)~d{\cal E}
\end{equation}
This probability can be integrated numerically.
\section{Summary and Discussion}
\noindent
We have presented here a power filter to search for gravitational wave signals
from burst sources in interferometric data.  The filter is designed to look
for signals of known duration and frequency bandwidth; when this is the only
available information, and the noise is stationary and Gaussian, the power
filter is optimal. Moreover, this filter is locally optimal for a wide class
of non-Gaussian noise, thus making it a useful tool to analyze real
interferometer data.

One shortcoming of the power filter is its inability to distinguish between
gravitational wave signals, and spurious instrumental artifacts which produce
time and band limited signals. This is mitigated by using multiple instruments
to detect gravitational wave bursts. As an added benefit, bursts identified as
noise can then be used for detector characterization. The extension of the
power filter to multiple instruments will appear in an upcoming
article\cite{pow_filt}; this article also contains a more complete discussion
of the filter, including implementation strategies and a comparison to matched
filtering.

In conclusion, we think that the power filter provides a useful tool for
gravitational wave data analysis.  It should play a significant role in
detector characterization for single interferometers, and should form the
basic building block in an hierarchical detection strategy using multiple
interferometers.

\acknowledgments{
This work was supported by the following NSF grants: PHY-9728704, PHY-9507740,
PHY-9970821, and PHY-9722189.}


\begin{thebibliography}{000}
\bibitem{Flanagan_E:1998}
\'{E}.~\'{E}. Flanagan and S.~A. Hughes, Phys. Rev. D \textbf{57}, 4535 (1998).
\bibitem{Thorne_K:1995}
K. S. Thorne, in \textit{Proceedings of the Snowmass 95 Summer Study on
Particle and Nuclear Astrophysics and Cosmology}, ed. E. W. Kolb and R. Peccei
(World Scientific, Singapore, 1995), and references therein.
\bibitem{filters}
M.~Feo, V.~Pierro, I.~M.~Pinto and M.~Ricciardi, in \textit{Edoardo Amaldi 
  Foundation Series Volume 2. Proceedings of the International Conference on 
  Gravitational Waves Sources and Detectors, March 19-23, 1996, Cascina (Pisa), 
  Italy}, ed. I. Ciufolini and F. Fidecaro, (World Scientific Publishing 
  Co., Singapore, 1997); N.~Arnaud, F.~Cavalier, M.~Davier, and P.~Hello, 
  Phys.\ Rev.\ D \textbf{59}, 082002 (1999); W.~G.~Anderson and 
  R.~Balasubramanian, Phys.\ Rev.\ D \textbf{60}, 102001-1 (1999); 
  S.~D.~Mohanty, ``A Robust Test for Detecting Non-Stationarity in Data from 
  Gravitational Wave Detectors'', gr-qc/9910027.
\bibitem{pow_filt}
W.~G.~Anderson, P.~R.~Brady, J.~D.~E.~Creighton and \'{E}.~\'{E}.~Flanagan, in
preparation.
\bibitem{likelihood}
L.~A.~Wainstein and V.~D.~Zubakov, \textit{Extraction of Signals from
   Noise} (Prenticehall, Engelewood Cliffs, New Jersey, 1962); L.~S.~Finn and
   D.~F.~Chernoff, Phys.Rev. D \textbf{47}, 2198, (1993).
\bibitem{Kassam_S:1988}
S. A. Kassam, \textit{Signal Detection in Non-Gaussian Noise}
  (Springer-Verlag, New York, 1988).


\end{thebibliography}
\end{document}